\documentclass[twocolumn,prb]{revtex4}
\usepackage{graphicx}
\usepackage{float}

\usepackage{color}

\begin{document}
\title{Crossover from 2-dimensional to 1-dimensional collective pinning in NbSe$_3$}

\author{E. Slot, H. S. J. van der Zant}
\affiliation{Department of NanoScience, Delft University of Technology, Lorentzweg 1, 2628 CJ Delft, The Netherlands}
\author{K. O'Neill, R. E. Thorne}
\affiliation{Laboratory of Atomic and Solid State Physics, Clark Hall, Cornell University, Ithaca, New York 14853-2501}
\date{\today}
\begin{abstract}
We have fabricated NbSe$_3$ structures with widths comparable to the Fukuyama-Lee-Rice phase-coherence length. For samples already in the 2-dimensional pinning limit, we observe a crossover from 2-dimensional to 1-dimensional collective pinning when the crystal width is less than 1.6~$\mu$m, corresponding to the phase-coherence length in this direction. Our results show that surface pinning is negligible in our samples, and provide a means to probe the dynamics of single domains giving access to a new regime in charge-density wave physics.
\end{abstract}

\pacs{PACS numbers: 72.15.Nj, 71.45.Lr, 73.23.-b}
\keywords{charge density wave; weak pinning; NbSe$_3$; Fukuyama Lee Rice; nanowire, mesoscopic}
\maketitle

Quasi-one-dimensional compounds comprised of metallic chains often undergo a phase transition to a charge-density wave (CDW) state.  The charge density is periodically modulated in space:\cite{peierls}
\begin{equation}
\rho(\vec{r})=\rho_0+\rho_1\cos[\vec{Q} \cdot \vec{r}+\phi(\vec{r})],
\end{equation}
where $\rho_1$ is the CDW amplitude and $\vec{Q}=2\vec{k}_F$ is the CDW wave vector, where $\vec{k}_F$ is the Fermi wavevector. The CDW phase~$\phi$ describes the local position of the CDW with respect to the underlying lattice.
Under the influence of an electric field the CDW can slide with respect to the lattice.\cite{frohlich} However, impurities and defects in the lattice pin the CDW so that depinning occurs only above a threshold electric field $E_T$.

Another consequence of pinning is that the CDW's phase $\phi$ varies on a characteristic length $l_{\phi}$, known as the Fukuyama-Lee-Rice length (and analogous to the Ovchinnikov-Larkin length in pinned vortex lattices.)  In clean CDW materials like NbSe$_3$ this length can be micrometers, allowing study of finite-size effects.\cite{ETsizeeffect}  Previous studies by McCarten~{\em et al.}\cite{McCarten} showed a crossover from 3-dimensional (3D) to 2-dimensional (2D) collective pinning in NbSe$_3$ when the crystal thickness (crystallographic a*-axis) was smaller than the phase correlation length $l_{\phi}$ in that direction. Finite-size effects play a key role in understanding CDW physics, since nearly all properties of pinned and moving CDWs depend on the dimensionality of CDW pinning and dynamics and thus on crystal size.

The width (crystallographic c-axis) of ribbon-shaped NbSe$_3$ crystals is almost always much larger than $l_{\phi}$ in this direction, so previous work has focussed on the 2D limit. Here, we have  prepared structures of varying widths both larger and smaller than $l_{\phi}$.  We observe a crossover from 2-dimensional to 1-dimensional (1D) pinning for the first time, and use this crossover to deduce the value of the CDW's phase-phase correlation length in the width direction. The 1D limit provides new opportunities for studying CDW physics, such as the effect of a single phase coherent domain on the dynamics of the CDW.\cite{mantel}

The depinning of a CDW is described by the model of Fukuyama, Lee and Rice~(FLR).\cite{fukuyama,lee} They consider the CDW as an elastic medium that deforms to adjust its phase in the presence of impurities.  The FLR Hamiltonian is
\begin{eqnarray}
\label{FLRH}
\mathcal{H}=\frac{1}{2}K \int d\vec{r}\left(\vec{\nabla }\phi \right)^{2}+~V_0\rho_1\sum_i \cos[\vec{Q} \cdot \vec{R_i}+\phi(\vec{R_i})] \nonumber \\
+\int d\vec{r}~\frac{\rho_{eff}\mathcal{E}\phi}{Q}.
\end{eqnarray}
The first term describes the elastic energy, where $K$ is the elastic force constant of the CDW; the second term describes the interaction with impurities located at $\vec{R_i}$, where $V_0$ is the impurity potential; and the third term describes the coupling of the CDW phase to an electric field~$\mathcal{E}$, where $\rho_{eff}$ is an effective condensed-charge density.

Although ``strong" impurities can locally pin the phase at low temperatures and require amplitude collapse for local phase motion,\cite{Larkin and Brazovskii} on larger scales the CDW is collectively pinned by elastic deformations of its phase on lengths~$l_{\phi}$ larger than the impurity spacing~$l_i=\sqrt[3]{1/n_i}$, where $n_i$ is the impurity concentration. These deformations allow the CDW to gain pinning energy by taking advantage of fluctuations in the interaction with randomly distributed impurities on that length scale. The elastic-energy cost of these deformations is balanced by the impurity-energy gain within a phase coherent volume. The elastic-energy cost has an upper limit, because $\nabla \phi$ does not exceed $\approx \pi /l_{\phi}$ within a phase coherent domain. For simplicity, we consider the case where $l_{\phi}$ is isotropic; the appropriate scaling for the anisotropic case is described in Refs.\ \onlinecite{McCarten} and~\onlinecite{lee}.

In 3D, the pinning-energy gain of a phase coherent volume is $-V_0\rho_1\sqrt{n_il_{\phi}^3}$. Minimizing the total energy (the elastic energy plus the pinning-energy gain) per unit volume of size $l_{\phi}^3$ yields
\begin{equation}
l_{\phi}={\left(\frac{2\pi^2K}{3V_0\rho_1\sqrt{n_i}}\right)}^2.
\end{equation}

The threshold electric field~$E_T$ is determined from the electrical energy needed to overcome the total energy per unit volume:
\begin{eqnarray}
\frac{\rho_{eff}E_T \phi_T}{Q}=V_0\rho_1\sqrt{\frac{n_i}{l_{\phi}^3}}-\frac{K}{2}\left(\frac{\pi}{l_{\phi}}\right)^2,
\end{eqnarray}
where $\phi_T$ is the angle through which the CDW has to evolve before it depins.

When the CDW is confined to a thickness $t$ less than the 3D phase correlation length in that direction, the CDW does not deform in that direction and the optimization of the pinning energy occurs in 2D, that is in the width and along the chains. The pinning energy changes to $-V_0\rho_1\sqrt{n_il_{\phi}^2t}$ and again minimizing the total energy results in a depinning field proportional to $1/t$. When the CDW is only confined in the width direction, $E_T$ has the same form with $t$ replaced by the width $w$.

When the CDW is confined in both thickness and width, phase deformations occur along the chains (1D), and the pinning energy becomes $-V_0\rho_1\sqrt{n_il_{\phi}wt}$ resulting in a threshold field proportional to $(1/wt)^{2/3}$.

These results assume that there is no surface pinning so that the elastic energy per unit volume is unchanged from the 3D case. Table~\ref{table:lphi_Et} summarizes  results for the phase correlation length $l_{\phi}$ and the depinning field~$E_T$ in the 3D, confined 2D and confined 1D cases. 

\begin{table}
\caption{\label{table:lphi_Et}The phase coherence length~$l_{\phi}$ and the threshold field~$E_T$ for weak pinning in case of no confinement (3D), confinement in the thickness~$t$ (2D) and for confinement in both width $w$ and thickness (1D). $l_{\phi}$ is given for the isotropic case. To compare $l_{\phi}$ to measurements, the anisotropy of the material should be taken into account. The appropriate scaling is described in Refs.~\onlinecite{McCarten} and~\onlinecite{lee}.}
\begin{tabular}{c c c}
\hline
\hline
\begin{tabular}{c}
confine-\\ment
\end{tabular}
&
$l_{\phi }$&
$E_{T}$\\
\hline \\[-3mm]
none&
$\left(\frac{2\pi^2K}{3V_0\rho_1\sqrt{n_i}}\right)^2$ &
$\frac{Q\left(V_0\rho_1\sqrt{n_i}\right)^4}{4\rho_{eff}\phi_T\left(\frac{2}{3} \pi^2  K\right)^3}$ \\[5mm]
thickness&
$\left(\frac{\pi^2 K}{V_0 \rho_1 \sqrt{n_i}} \right) \sqrt{t}$ & 
$\frac{Q\left(V_0\rho_1\sqrt{n_i}\right)^2}{2\rho_{eff}\phi_T\left(\pi^2  K\right) } \hspace{1em}\frac{1}{t}$ \\[5mm]
\begin{tabular}{c}
width-\\
thickness
\end{tabular}&
$\left(\frac{2\pi^2K}{V_0\rho_1\sqrt{n_i}}\right)^{2/3}\sqrt[3]{wt}$ &
\hspace{1em}$\frac{Q\left(V_0\rho_1\sqrt{n_i}\right)^{4/3}}{2\rho_{eff}\phi_T\left(2\pi^2 K \right)^{1/3}} \left(\frac{1}{wt}\right)^{2/3}$ \\
\hline
\hline
\end{tabular}
\end{table}

\medskip
To probe width-dependent pinning and the 2D to 1D crossover we have fabricated small NbSe$_3$ structures using two techniques. In the first, a crystal is placed on top of a gold contact pattern. The crystal width is then reduced by using either a focused-ion beam (FIB) or by reactive-ion (SF$_6$) etching. The crystal width can be controllably reduced to $\sim $~200~nm, provided that the thickness is comparable to or smaller than the final width.  Measurements are then performed in a 4-point configuration. A full description of the fabrication process can be found in Refs.\ \onlinecite{Kevin Ecrys} and~\onlinecite{Erwin Ecrys}. In all cases, the initial crystals were thinner than the corresponding phase correlation length so that the pinning was initially in the 2D limit.

The second technique does not use etching to reduce the crystal size. Instead, bulk NbSe$_3$ crystals are mixed into pyridine and then shaken in an ultrasonic bath. The ultrasound cleaves the crystals, producing a suspension of NbSe$_3$ nanowires with typical cross sections $(t \times w)$ of 50~nm$\times$100~nm and lengths of 20~$\mu $m. A drop of this suspension is put on a Si substrate coated with insulating SiO$_2$ and left to evaporate. The position of the nanowires is determined with respect to predefined makers on the substrate, and e-beam lithography followed by Ti and Au depositions are used to define a contact pattern on top of the wires. Both 2-point and 4-point structures are made with typical contact separations of 2 to 5~$\mu $m. The two fabrication techniques complement each other, producing samples with cross sections varying by nearly 5 orders of magnitude from 500~nm$^2$ to 20~$\mu$m$^2$.

The depinning field $E_T$ was determined from differential resistance (d$V$/d$I$) measurements in a helium flow cryostat using a standard lock-in amplifier technique.  To characterize current homogeneity within the samples, we have applied large amplitude ac signals to mode lock the internal CDW ``washboard" frequency to multiples of the applied frequency.  We have observed complete mode locking on samples prepared by both fabrication techniques, indicating homogeneous current flow and that the fabrication techniques do not reduce the high quality of the crystals. From the distance between mode-locked steps, we have determined the sample cross sectional areas and these agree to within 10\% with cross sections determined from the room-temperature resistance.  Similar results are obtained by comparing the 2- and 4-point resistances, indicating that contact resistances are less than 10\% of the nanowire resistance. Data presented here were obtained at $T=120$~K, where contributions to the measured voltage due to the current conversion process at current contacts are minimal.

\medskip
\begin{figure}[t]
\includegraphics*[angle=0,width=8.6cm]{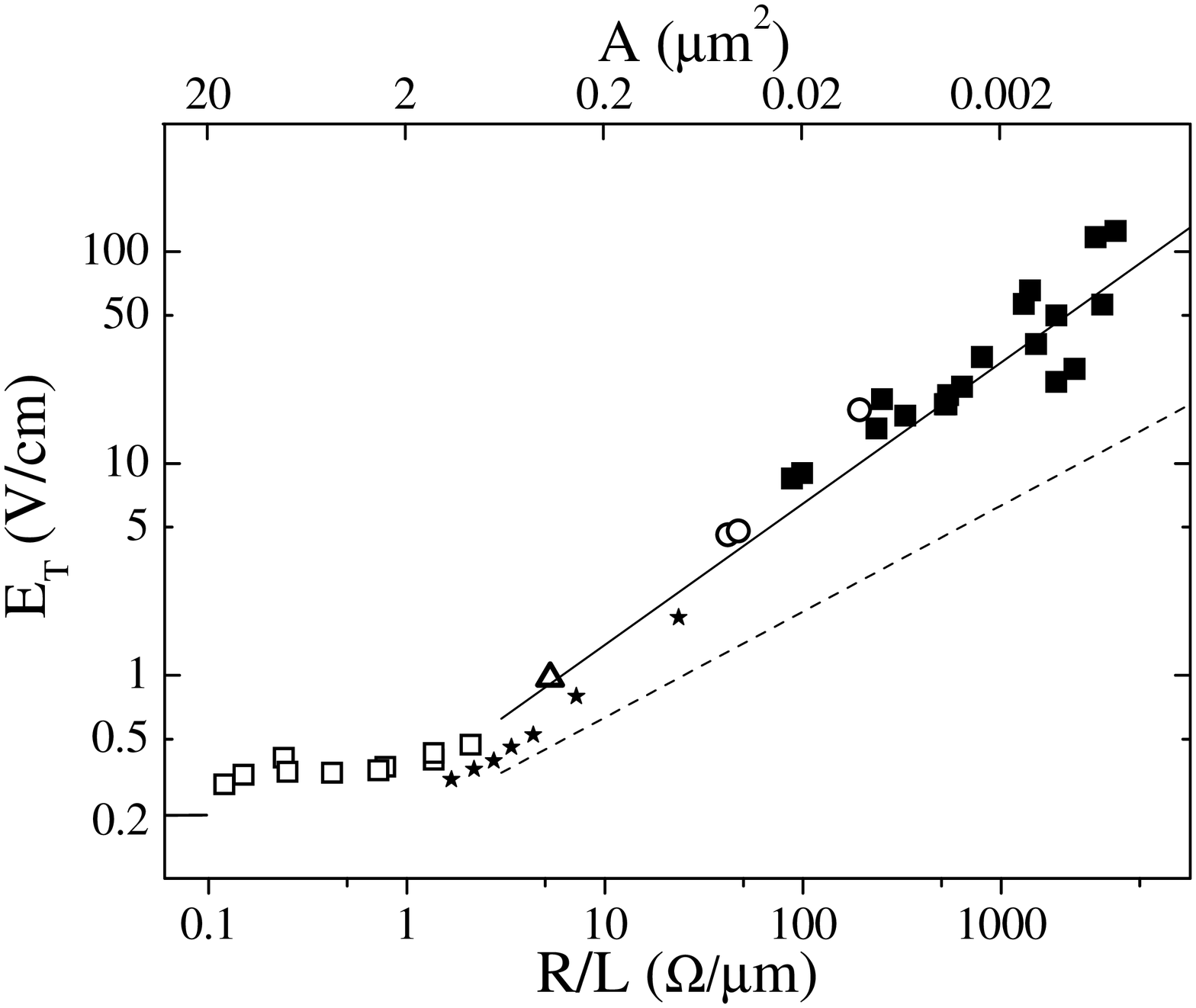}
\caption{\label{ETvsRL}The threshold field $E_T$ at $T=120$~K as a function of the room-temperature resistance~$R$ over the voltage probe separation~$L$. The cross section $A=wt$ for rectangular cross sections is inversely proportional to $R/L$. The open symbols are data of the three FIB processed crystals, with thicknesses of (squares) 0.56~$\mu $m, (triangle) 0.58~$\mu $m, and (circles) 0.20~$\mu $m. The stars are data of an SF$_6$ etched crystal with a thickness of 0.7~$\mu $m. These are all smaller than the lower bound of the bulk (3D) CDW phase-phase correlation length in the thickness direction (a*-axis) of 0.80~$\mu $m measured by X-ray diffraction.\cite{xray} For reference, 0.2~V/cm is the threshold field of an unprocessed sample with $t=0.5~\mu$m, taken from Ref.~\onlinecite{McCarten}. Each closed square represents the data of a different nanowire. The solid line displays the $A^{-2/3}=(R/L)^{2/3}$ dependence expected for 1D pinning. The dashed line displays the $A^{-1/2}=(R/L)^{1/2}$ dependence expected for surface pinning.}
\end{figure}

Figure~\ref{ETvsRL} shows the threshold field~$E_T$ as a function of $R/L$ for the FIB and SF$_6$ etched samples as well as for the much smaller sonicated samples.  $R/L$ is the room-temperature resistance between voltage probes, and with the good assumption that the room-temperature resistivity is independent of sample size, $R/L$ is inversely proportional to the cross section $A=wt$. When $R/L$ is less than~$\approx$~1~$\Omega /\mu $m corresponding to a cross-sectional area larger than 2~$\mu $m$^2$, $E_T$ does not depend on the cross section.  For $R/L>1~\Omega /\mu$m, $E_T$ increases strongly with decreasing cross section. On the log-log plot of Fig.\ \ref{ETvsRL}, $E_T$ follows the solid line with a slope of 2/3 over more than two decades.

\begin{figure}[b]
\includegraphics*[angle=0,width=8.6cm]{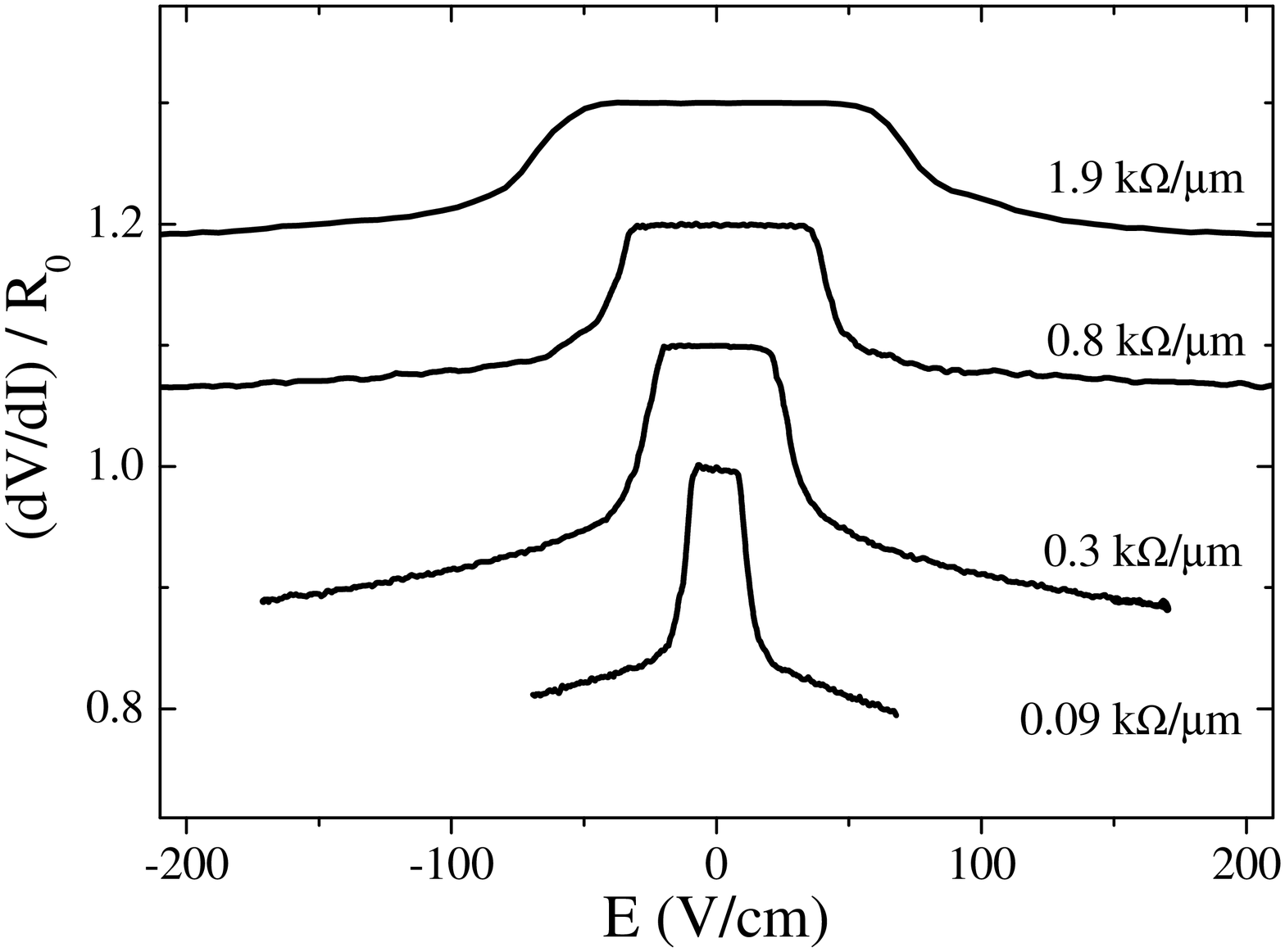}
\caption{\label{dVdIs}The differential resistance d$V$/d$I$ normalized by the zero-bias resistance $R_0$ for four nanowires as a function of electric field at $T=120$~K. The curves have offsets of 0, 0.1, 0.2, and 0.3. The room-temperature resistance per unit length $R/L$ is shown next to each curve. The onset of CDW conduction becomes more rounded for higher $R/L$ values.}
\end{figure}

Figure~\ref{dVdIs} shows the d$V$/d$I$ curves for four nanowires. The $R/L$ values are shown next to each curve. For low $R/L$ values, the threshold for sliding is easy to determine visually by a sharp decrease of the differential resistance. As the cross section decreases the threshold for sliding is more gradual and harder to determine due to thermal rounding.\cite{McCarten,rounding} This contributes to the increased scatter in the data of Fig.\ \ref{ETvsRL} for very small crystals ($R/L>1$~k$\Omega /\mu$m). For nanowires with $R/L>10$~k$\Omega$/$\mu$m, the threshold for sliding could not be determined at $T=120$~K.

We have also investigated $E_T$ at temperatures of 130~K to 100~K. We have obtained plots where $E_T$ increases with $R/L$ following a straight line on a log-log scale similar to Fig.\ \ref{ETvsRL}. The exponents at $T=130$~K and $T=120$~K are 0.67, and increase when decreasing temperature further to a value of 0.76 at $T=100$~K. This increase may be due to larger contributions to $E_T$ from phase slip processes at lower temperatures.

\medskip
What is the origin of $E_T$'s size dependence in small crystals?  The CDW in the crystal larger than 2~$\mu $m$^2$ in Fig.\ \ref{ETvsRL} begins in the 2D collective pinning regime where $E_T \propto 1/t$, and where pinning related CDW deformations are cut off by the crystal thickness.  The crossover to a width-dependent $E_T$ as the crystal is etched to small widths is due to a crossover from 2D to 1D collective pinning, in which transverse CDW deformations are cut off by both the thickness and width.  This is strongly supported by the measured exponent $E_T \propto A^{-2/3}$ for crystals ranging over three orders of magnitude in cross section, which corresponds with the prediction for 1D collective pinning in Tab.\ \ref{table:lphi_Et}. The crossover width depends on the thickness of the crystal as $\sqrt{t}$ and is smaller for thinner crystals.\cite{comment} The crossover width of the sample represented by the squares in Fig.\ \ref{ETvsRL} is 1.6~$\pm $~0.2~$\mu $m. This value is a factor of two larger than the bulk (3D) CDW correlation length in the c*-axis of 0.75~$\mu $m measured by X-ray diffraction.\cite{xray} This provides strong additional support for a 2D to 1D pinning origin.

An increase of $E_T$ with decreasing crystal cross section could also arise due to pinning by crystal surfaces.\cite{ETsizeeffect} In this case $E_T$ is determined by the crystal surface-to-volume ratio.  For thin but wide ($w\gg 1$~$\mu $m) crystals, $E_T \propto 1/t$ for both surface pinning and 2D collective pinning, making these mechanisms difficult to distinguish.\cite{McCarten} But in crystals where the CDW is confined in both width and thickness, $E_T$ due to surface pinning is proportional to $(w+t)/wt$. Since the width-to-thickness ratio $w/t$ shows only a small sample-to-sample variation compared with the three orders of magnitude variation of cross-sectional area in Fig.\ \ref{ETvsRL}, this simplifies to $E_T \propto A^{-1/2}$. This differs from the 1D collective pinning prediction $E_T \propto A^{-2/3}$.  As indicated by the dashed line in Fig.\ \ref{ETvsRL}, $E_T \propto A^{-1/2}$ is clearly inconsistent with experiment. The results for $E_T$ with $R/L>1$~$\Omega /\mu $m clearly show that surface pinning can be excluded.

In the 1D limit, the total energy of a phase coherent domain is proportional to (wt)$^{2/3}$. Accounting for NbSe$_3$'s anisotropy and using $K=3.5$~meV/\AA, $V_0\rho_1=4$~meV and $n_i=2.5\times 10^{16}$~cm$^{-3}$ taken from Refs.\ \onlinecite{McCarten} and~\onlinecite{serge impurity}, we estimate the total energy of a 1D domain in a crystal with $R/L=2$~k$\Omega /\mu$m to be only $\approx 4.5$~kT at $T=120$~K, compared with bulk 3D values of $10^5$~K.  This small energy explains the pronounced thermal rounding in the depinning transition visible in the d$V$/d$I$ for these small crystals in Fig.\ \ref{dVdIs}.

The number of pinning centers per unit length decreases with decreasing cross-sectional area. Eventually the separation between pinning centers is comparable to the phase-coherence length along the length. This changes pinning for samples with small cross-sectional areas. The separation of pinning centers for our smallest sample is estimated to be 100~nm (using $n_i=2.5~\cdot~10^{16}$~cm$^{-3}$ and $wt=400$~nm$^2$), while the predicted 1D phase-coherence length, taken from Table~\ref{table:lphi_Et}, is 350~nm for the smallest sample. This means that for all samples shown in Fig.~\ref{ETvsRL} the phase-coherence length is larger than the separation between pinning centers. Thus the weak pinning limit applies to all our samples.

In summary, we have shown a dimensionality crossover from 2-dimensional to 1-dimensional weak pinning in NbSe$_3$. 1D pinning is observed when both the width and thickness are smaller than the CDW's bulk phase coherence length in these directions. The observation of 1D pinning behavior is not an artefact of the fabrication technique, since it is observed for samples prepared using three different methods.  From the width dependence of $E_T$ we estimate the CDW's phase correlation length in that direction to be 1.6~$\mu $m, comparable to the value obtained by X-ray diffraction.\cite{xray}

\medskip
The ability to reach the 1D limit should provide new opportunities for studying CDW physics.  For example, in ordinary size crystals the cross section contains a large number of phase correlated domains in the pinned state and of dynamically correlated domains in the depinned state.  In 1D crystals, the cross section will contain a single domain in both the pinned state and in the depinned state at modest fields, and the length of the domain will be $\approx $~10~$\mu$m.  This will allow the detailed dynamics on the scale of a single domain to be explored with nanofabricated probes spaced along the crystal.\cite{mantel, adelman 1995 prb} Large-scale numerical simulations by Matsukawa\cite{matsukawa} suggest that the CDW's dynamical correlation length should be larger or comparable to the pinned correlation length out to at least a few times $E_T$.  Theoretically, the size of the critical regime near the depinning transition should be larger in 1D,\cite{fisher 1983, middleton, myers} and may finally make it experimentally accessible, although equally interesting finite-size effects in the dynamics\cite{middleton, myers} should also be more important.
\acknowledgements{We appreciate useful discussions with P.\ H.\ Kes and we thank S.\ V.\ Za\u{\i}tsev-Zotov and M.\ A.\ Holst for work on the nanowires. This work was supported by the Dutch Foundation for Fundamental research on Matter (FOM), the Netherlands Organisation for Scientific Research (NWO), INTAS (Project 01-0474) and the National Science Foundation (NSF) (DMR 0101574 and INT 9812326). K.\ O'Neill was supported by a U.\ S.\ Department of Education Fellowship. Nanofabrication work was performed at DIMES in Delft and at the Cornell Nano-Scale Science \& Technology Facility, supported by the NSF (ECS-9731293).}

\end{document}